# Metasurface for programmable quantum algorithms with quantum and classical light

Randy Stefan Tanuwijaya[1], Hong Liang[1], Jiawei Xi[1], Tsz Kit Yung[1],
Wing Yim Tam[1], and Jensen Li[1*]

[1]Department of Physics, The Hong Kong University of Science and Technology,

Clear Water Bay, Kowloon, Hong Kong, 999077, P. R. China

**Abstract**

Metasurfaces have recently opened up applications in the quantum regime, including quantum tomography and the generation of quantum entangled states. With their capability to store a vast amount of information by utilizing the various geometric degrees of freedom of nanostructures, metasurfaces are expected to be useful for processing quantum information. In this study, we propose and experimentally demonstrate a programmable metasurface capable of performing quantum algorithms using both classical light and quantum light at the single photon level. Our approach encodes multiple programmable quantum algorithms, such as Grover's algorithm and the quantum Fourier transform, onto the same metalens array on a metasurface. A spatial light modulator selectively excites different sets of metalenses to carry out the quantum algorithms, while the photon arrival data or interference patterns captured by a single photon camera are used to extract information about the output state. Our programmable quantum metasurface approach holds potential as a cost-effective means of miniaturizing components for quantum computing and information processing.

* email: jensenli@ust.hk

While superconducting circuits and trapped ions are two promising physical technologies for implementing quantum computing, photonic implementation remains a versatile platform [1-3]. Photonic implementation not only allows for the investigation of fundamental questions in quantum mechanics [4-6], but also provides a practical way to demonstrate the advantages of quantum algorithms in solving problems that are significantly slower on classical computers [7-9]. In fact, classical optics already possesses key elements of quantum computing, such as superposition, interference, and to a large extent, non-separable states analogous to entanglement without quantum non-locality [10,11]. By utilizing the different internal degrees of freedom of a single photon, it becomes possible to construct non-separable states and perform quantum algorithms at both the single-photon level and even using classical light [12,13].

Tremendous efforts have been devoted to simulating various quantum algorithms using classical light as an easily accessible testbed. For instance, quantum algorithms involving up to two-qubit operations can be constructed using common optical components, such as beam splitters, polarization beam splitters, lenses, and Dove prisms, by leveraging the two paths and polarization degrees of freedom [12,13]. Recent experiments have explored the use of vortex beams generated or manipulated by a spatial light modulator (SLM) to achieve quantum algorithms with up to 4 qubits, utilizing 16 orbital angular momenta (OAM) [14,15]. Additionally, more efficient and programmable utilization of the SLM has demonstrated the implementation of the Deutsch-Jozsa algorithm with an effective complexity of 20 qubits [16,17]. These principles can also be applied at the single-photon level when using a quantum source, where practical demonstrations often employ the heralding technique to enhance detection purity and signal-to-noise ratio [17,18]. Moreover, by working with photon counting in the quantum regime, it becomes possible to investigate two-photon physics within the same linear optical system. This approach opens up avenues for studying boson sampling, a non-universal quantum computation method that is useful for evaluating the permanents of matrices [19].

On the other hand, metasurfaces, which consist of a single layer of nanostructures, have proven to be highly useful in multiplexing or demultiplexing optical information. They have been applied successfully in generating vortex beams and holograms with fine resolution [20-24]. As we move towards the quantum regime, we are witnessing early demonstrations of metasurfaces in quantum tomography, the generation of high-dimensional entanglement, quantum imaging, and the manipulation of two-photon interference [25-32]. Furthermore, due to their capability to store a significant amount of information regarding light-matter interaction, metasurfaces offer potential for processing quantum optical information and implementing quantum algorithms. Some studies have even proposed the use of metamaterials to simulate quantum algorithms in the classical domain [33-35].

In this work, we have developed a platform that utilizes a geometric-phase metalens array on a metasurface to execute quantum algorithms, enabling the implementation of unitary transformations in a single step. By employing a spatial light modulator (SLM) and a single photon camera for the generation and detection of input and output states, we have successfully performed programmable quantum algorithms in the optical domain. This includes the implementation of Grover's search (GS) and the quantum Fourier transform (QFT) by selectively illuminating the metalenses. To validate our approach, we conducted simulations and experiments using both classical and quantum light. The programmable nature of our approach offers the potential for miniaturizing quantum optical circuits, providing a low-cost and integrated solution for quantum information processing.

**Results and discussion**

**Computing scheme using metalens array for quantum algorithms**

Our goal is to implement a specific quantum algorithm for *n*-qubits, which can be summarized as a transformation: $|j\rangle \to \sum_{j=0}^{2^n-1} U_{ij} |i\rangle$, where $U$ is a unitary matrix, using a metasurface. The integer $i$ or $j$ within the "ket" notation ranges from 0 to $2^n - 1$, representing the $n$-qubit basis for either the input or output state. Our metasurface approach is shown in Fig. 1a. A spatial light modulator (SLM) is utilized to prepare the input state $\sum_{j=0}^{2^n-1} \psi_j^{(\text{in})} |j\rangle$, where $\psi_j^{(\text{in})}$ represents the complex electric field normally incident on the lens located at $R_j$. The SLM operates in vertical polarization, which is subsequently converted to left-circular polarization (LCP) before illuminating the metasurface. Our metasurface comprises 9 metalenses, out of which four are employed to perform a specific 2-qubit ($n = 2$) unitary operation. Each metalens at $R_j$ diffracts the incident electric field $\psi_j^{(\text{in})}$ into a discrete set of output directions $\boldsymbol{k}_i$, with the designed complex amplitudes $U_{ij}$ (different colors, although physically all at the same wavelength, illustrate different output directions; for instance, a green beam corresponds to the normal exit direction, while a cyan beam corresponds to a $10°$ deflection to the left). In

our case, each metalens converts LCP to right-circular polarization (RCP) using a custom complex transmission amplitude and phase profile given by the expression:

$$t_j(\boldsymbol{\rho}) = \sum_{i=0}^{2^n-1} U_{ij} \exp(i\, \boldsymbol{k_i} \cdot \boldsymbol{\rho}), \qquad (1)$$

where $\boldsymbol{\rho}$ denotes the transverse position measured from the center of the entire metasurface. Consequently, light from different metalenses with the same output direction interferes in the far-field regime, resulting in an electric field $\psi_i^{(\text{out})}$ in RCP. The interference pattern $\left|\psi_i^{(\text{out})}\right|^2$ around each direction $\boldsymbol{k_i}$ is captured by a single photon camera. The entire unitary operation can be summarized as $\psi_i^{(\text{out})} = \sum_{j=0}^{2^n-1} U_{ij} \psi_j^{(\text{in})}$ in terms of the complex field amplitudes.

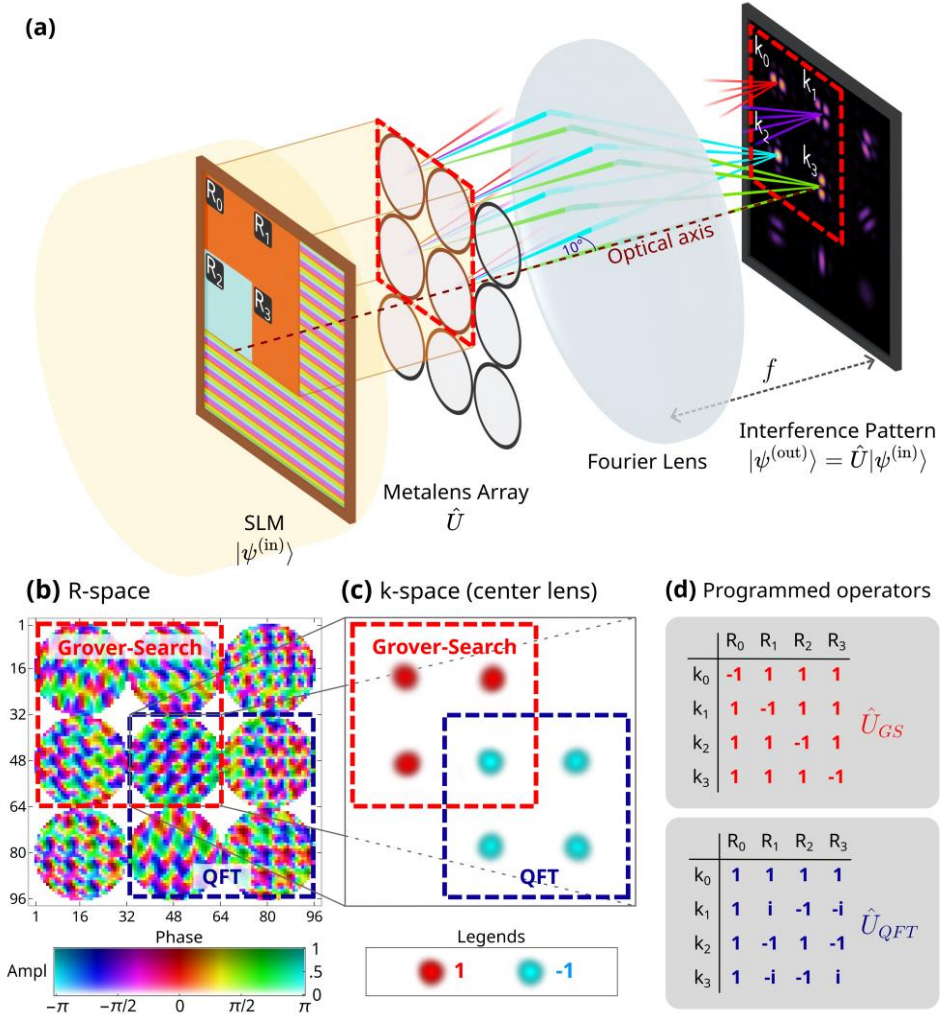

**Fig. 1 | Computing metasurface for programmable quantum algorithm. a,** Overall schematic of the computing metasurface for implementing two-qubit unitary quantum operations. A spatial light modulator (SLM) selectively excites the metalens array and prepares the two-qubit input state $|\psi^{(\text{in})}\rangle$ by modulating the complex incident E-field for

each selected metalens at $R_0$ to $R_3$. Each metalens diffracts the beam into multiple output directions, with amplitude and phase modulated by the $U$-matrix. The interference patterns of the diffracted beams in the same direction are projected onto the Fourier plane, where the intensity profile around $k_0$ to $k_3$ yields the output state $|\psi^{(\text{out})}\rangle$. **b,** Complex transmission amplitude and phase profile of the metalens array, encoding both Grover's Search (GS) and Quantum Fourier Transform (QFT). **c,** Far-field intensity profile when only the center lens is excited to implement a column of the unitary $U$ matrix. The color shows phase modulation for different output directions (all of the same wavelength). **d,** The corresponding matrices for the two programmed quantum operators, $\widehat{U}_{GS}$ and $\widehat{U}_{QFT}$. Programmability is achieved by selecting the metalens excitation ($R_0$ to $R_3$) and the projected interference patterns ($k_0$ to $k_3$). For example, selecting red square in (**b**) and (**c**) enables GS, while selecting blue square enables QFT.

Our scheme incorporates programmability using a static metasurface with shared resources for different quantum algorithm choices through selective illumination. The input field $\psi_j^{(\text{in})}$ on a specific metalens at $R_j$ is prepared by the corresponding region of a (phase) SLM at normal incidence. A certain fraction of these pixels is deactivated by introducing a phase gradient, diverting the light at an angle that does not fall on the lens array. This process sets $\psi_j^{(\text{in})}$ with both amplitude and phase [31]. The metalens then implements the *j*-th column of the unitary matrix $U$ through a transmission profile at each lens. Fig. 1b illustrates the transmission profile (amplitude and phase in a color map) of our metasurface, comprising a total of 9 metalenses. It should be noted that an additional quadratic focusing phase profile is added to each transmission profile, integrating the schematic Fourier lens shown in Fig. 1a onto the metalens array. Our metalens array is programmed with two different quantum operations: Grover's search (GS) or quantum Fourier transform (QFT), achieved by illuminating an arbitrary 2-qubit input state on the 4 upper left (dashed red square) or lower right lenses (dashed blue square) in Fig. 1b, referred to as R-space, using the SLM. The lens labels $R_0$ to $R_3$ for the GS are indicated in Fig. 1a, while same lens labels are defined in a similar row-major order for the lenses used in QFT. On the other hand, the four output directions $k_0$ to $k_3$ for the GS or QFT are shown in Fig. 1c, referred to as k-space, again using dashed red square or blue square. For convenience, we have adopted the same label order as the convention used for R-space, but other choices are possible as long as 4 out of the 9 directions are selected as outputs in defining the $U$ matrix. The corresponding unitary matrices of GS ($U_{GS}$) and QFT ($U_{QFT}$) are shown Fig. 1d (normalization factor omitted). Taking the center lens, $R_3$, as an example, it generates $\{1,1,1,-1\}$ at its 4 output directions for $U_{GS}$, or $\{-1,-1,-1,-1\}$ (for the same lens now labeled as $R_0$) at its 4 output directions for $U_{QFT}$ (a global $\pi$ phase is added to the entire matrix without altering unitarity). Thus, $R_3 \to k_3$ for $U_{GS}$ and $R_0 \to k_0$ for $U_{QFT}$ both implement the same matrix element $-1$, indicating that this resource is shared between the two algorithms. These 8 matrix elements, with one of them shared and implemented by the center lens, are shown in Fig. 1c. The sharing of resources allows for a more compact implementation of the metalens array and the incorporation of additional functions given limited resources. For the entire (9 lenses) × (9 directions) matrix, there are unused matrix elements where we can embed more operations (for further details, refer to the Supplementary Materials).

**Geometric-phase metasurface and experimental setup**

Fig. 2a showcases the Scanning Electron Microscope (SEM) image of the metasurface, with a close-up view provided in Fig. 2b. The design consists of nine metalenses, each comprising 32 unit cells along the diameter (approximately 20μm in diameter). Each metalens is composed of nano-slots with variable orientations. In particular, each unit cell, with a periodicity of $p = 620$nm, contains two pairs of rotated nanoslots (195 nm long, 50 nm wide) with rotation angles $\theta_1$ and $\theta_2$, as depicted in Fig. 2c. The chosen periodicity is smaller than the operational wavelength of $\lambda = 810$nm to suppress higher-order diffraction. Regarding the two pairs of rotating slots within a unit cell, we focus on achieving cross-circular polarization (CP) from LCP to RCP in our scheme, contributing to a transmission with an amplitude proportional to $|\cos(\theta_1 - \theta_2)|$ and an overall geometric phase of $\theta_1 + \theta_2$ [36]. By selecting appropriate values of $\theta_1$ and $\theta_2$ at each unit cell within the metalens at $R_j$, the resultant complex transmission amplitude profile is designed to match the specifications outlined in Eq. (1), where now the $U$ matrix represents the whole $9 \times 9$ operation matrix obtained by combining all the unitary matrices of the programmed quantum algorithms. Here, $i$ ranges from 0 to 8, corresponding to the 9 output directions from the same lens at $R_j$. The design of metalens array, incorporating the nanoslots with the orientation profile, is fabricated on silver-coated glass (50nm coating) using the focused ion beam (FIB) technique.

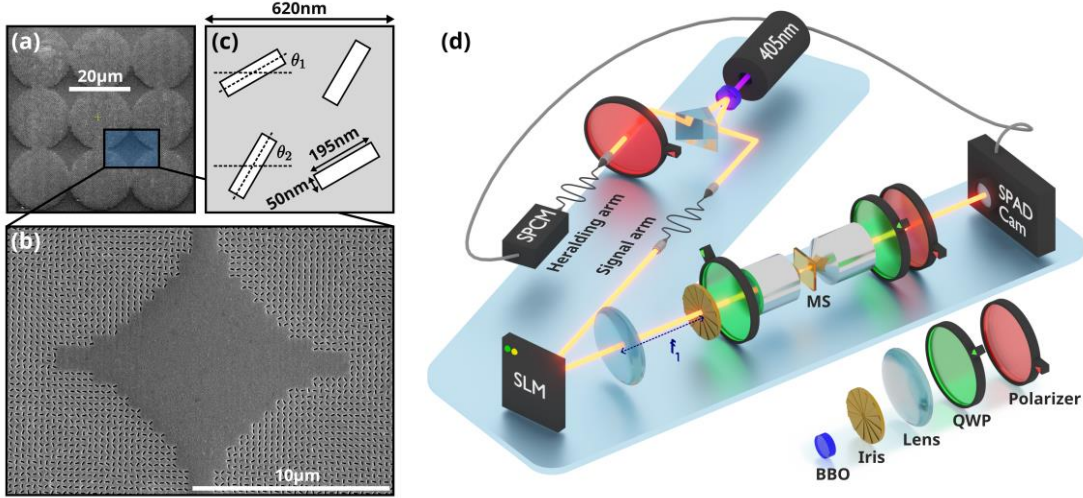

**Fig. 2 | Geometric-phase metalens array and heralding photon experimental setup. a,** SEM image of the metasurface designed for operational wavelength λ = 810 nm. **b,** Enlarged SEM view of the metalens array showcasing the quality of the metalens. **c,** Unit cell for controlling amplitude and phase response, composed of two pairs of rotated nano-air-slots on a silver film (50nm thick) coated glass substrate. **d,** Experimental setup using a heralding photon source for imaging the far-field interference patterns resulting from the quantum operation.

The experimental setup, incorporating an entangled quantum light source, is illustrated in Fig. 2d. A 405nm laser (CrystaLaser DL-405-400) pumps a type-II β-barium borate (BBO) crystal to generate two photons at a wavelength of 810nm in a polarization-entangled state of $1/\sqrt{2}\left(|HV\rangle+|VH\rangle\right)$, with one photon sent to the heralding arm and the other to the signal arm. A polarizer on the heralding arm selectively filters horizontally (H) polarized photons, which are then detected by a Single Photon Counting Module (SPCM) triggering the Single Photon Avalanche Diode (SPAD) camera (Pi Imaging SPAD 512S) located at the end of the signal arm. The SPAD camera operates in gated mode with a 16-ns coincidence window. The heralding process only triggers imaging for those photons with vertical (V) polarization that arrive at the Spatial Light Modulator (SLM) (Holoeye Pluto 2.1) in the signal arm. The SLM generates the input state and selectively excites the metalens. As mentioned earlier, to control the amplitude from the phase-SLM, some light is diverted and blocked by an iris positioned at the focal plane of a lens with $f_1 = 40$cm. A pair of orthogonal Quarter Wave Plates (QWP) is placed before and after the metasurface to convert the linearly polarized incident beam to circular polarization (CP) and vice versa. The first QWP, with a fast axis oriented at the anti-diagonal, converts V-polarized light to LCP. Upon interacting with the metasurface, the photons diffract into different output directions. The second QWP, with a fast axis oriented at the diagonal, and an H-polarizer are used to convert the polarization back to linear polarization that is orthogonal to the input light in the signal arm. Consequently, interference patterns corresponding to different output directions are formed at the focal plane of the metalens array and captured by the SPAD camera. Extended exposure time (a total of 20,000 frames with each frame lasting 0.3s) is utilized for data acquisition due to the extremely low photon count from the quantum source, even with the use of the heralding technique to increase the signal-to-noise ratio for detection.

**Experimental characterization**

We first examine the operations of the metasurface using classical light. In this case, we replace the part before the SLM in Fig. 2d with a 810 nm laser (OBIS LX 808 nm 150 mW Laser). The utilization of classical light for excitation effectively enhances the signal, resulting in shorter acquisition times (5-10ms). It also leads to higher resolution in the interference patterns formed at the focal plane of the metalens array. This is because the higher receiving power can be distributed among more pixels on the single photon camera. These investigations facilitate the more challenging task of extracting information from the interference patterns in the subsequent stage with a quantum light source.

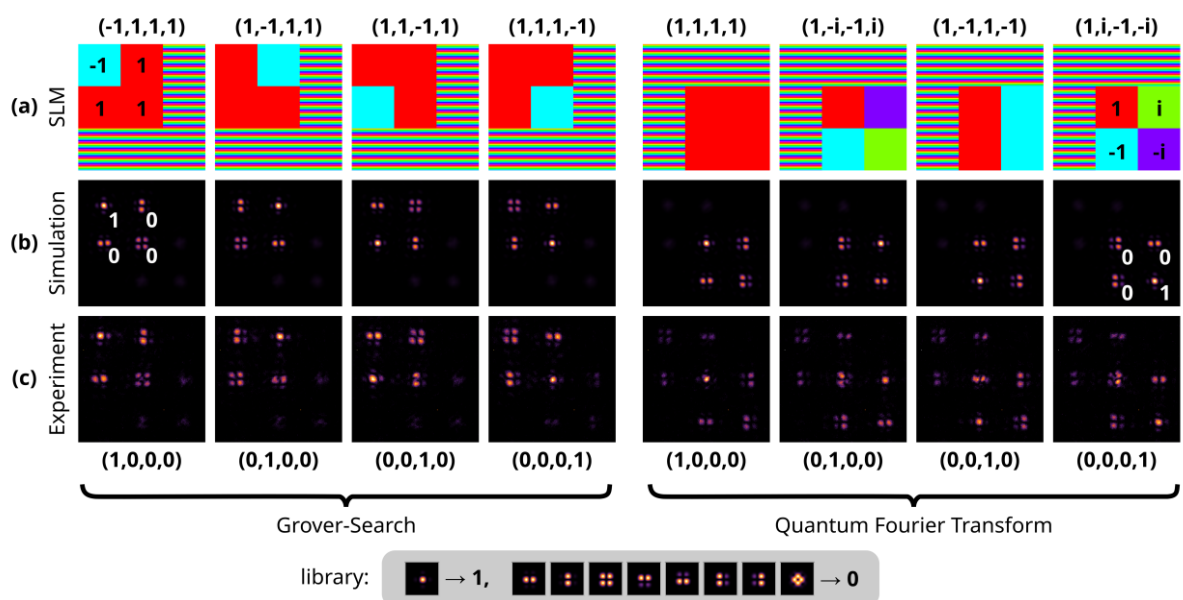

**Fig. 3 | Simulation and experimental results using classical light source for the Grover-Search (left) and QFT (right). a,** SLM phase profile for each input state for the two algorithms: four input states for each algorithm listed at the top of each column. Color

code: red for 1, cyan for -1, green for $i$, and purple for $-i$, shaded regions (phase gradient) are effectively inactive. **b,** Numerical simulation of the far-field interference pattern. **c,** Experimental result using a classical light source. The input and output states are labeled at the top and bottom of each column. The library for decoding the interference pattern into the output state for the representative cases of value 0 and 1 is listed in the bottom inset.

Fig. 3 displays the simulation and experimental results obtained using a classical light source for two quantum algorithms: GS (left 4 columns) and QFT (right 4 columns). For each algorithm, we test it using four different input states, which are labeled at the top of each corresponding result column. Let us begin with the first column, which corresponds to the input state $(-1,1,1,1)$ for the GS algorithm. Our objective is to pick the basis with $\pi$ phase shift and output the state $(1,0,0,0)$ with unit probability on the corresponding basis using the metalens array to implement the unitary matrix $U_{GS}$ (as shown in the matrix multiplication in Fig. 1d). The input state is initially encoded as the phase profile on the SLM (Fig. 3a) using the four upper-left lenses in a row-major manner: a region marked as -1 has a constant SLM phase of $\pi$ (cyan color), while a region marked as 1 has a constant SLM phase of 0 (red color). Only the corresponding metalenses are selectively excited to implement the GS algorithm, while the other regions display a phase gradient to deflect the light away from the metalens array. In the far field, the interference patterns are formed at the output directions, as shown in the numerical simulation (by propagating the complex field to the focal plane of the metalens) in Fig. 3b and the experimental results captured by the single photon camera in Fig. 3c. The interference patterns (intensity) for the two results closely resemble each other and represent the output state in the upper left region, indexed also in a row-major manner. The intensity $\left|\psi_i^{(out)}\right|^2$ at direction $k_i$ can then be determined from the center of each interference pattern. It is worth noting that only the upper left 4 directions exhibit notable interference patterns (as seen Fig. 1c). To achieve higher accuracy, we extract the intensity using the entire interference pattern. Some representative cases from the library are shown at the bottom of Fig. 3, illustrating the different possible patterns for value 0 (fully destructive, resulting in dark centers) and 1 (fully constructive, resulting in a bright center). By comparing the interference patterns with the library, we find that the output state agrees to the expected result $(1,0,0,0)$ for the GS algorithm. In the meantime, for the sake of convenience of our discussion, we discretize the output state to be either fully destructive or fully constructive interference. More details are given in Supplementary Materials regarding the decoding process for the full spectrum of intensity values, other than 0 and 1, which will be used in the next step.

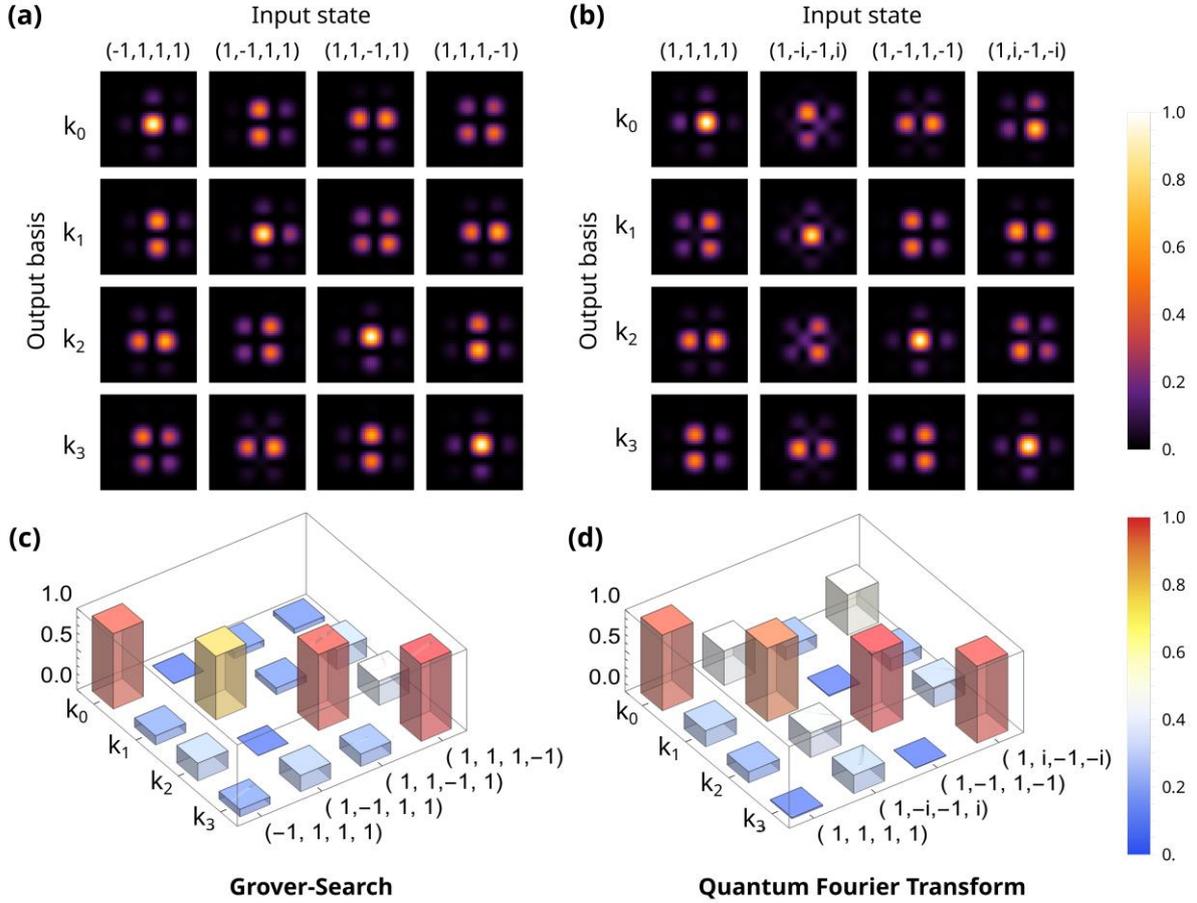

**Fig. 4 | Experimental results using quantum heralding photon source. (a-b),** Interference patterns after up-sampling for the GS (left) and QFT (right) algorithms. **(c-d),** Normalized fitted output state amplitudes extracted from the interference patterns for the corresponding algorithms.

For completeness, the results for the other three different input states with $\pi$ phase change at the corresponding input basis, are shown in the second to forth columns. The simulation, experiment results exhibit good agreement, while the decoded output states are shown below the experimental results. To test the quantum Fourier transform (QFT) using the same technique, we select the input states to be $(1,1,1,1)$, $(1,-i,-1,i)$, $(1,-1,1,-1)$ and $(1,i,-1,-i)$ in order to detect the different "frequencies" in the state. The corresponding phases are set by the SLM, as shown in the first row of Fig. 3a. The output states correspond to the columns of the $4 \times 4$ identity matrix. Notably, only the four lower-right output directions exhibit significant interference patterns. The simulation and experimental interference patterns are in agreement and can be decoded (in a row-major manner) to yield the expected output states.

With the previous experience gained from using classical light to confirm the metasurface's functionality, we now shift our focus to discussing the results obtained with a quantum light source. This presents an additional challenge in extracting information due to the lower number of photons reaching the single photon camera. In this case, we can only use a significantly smaller number of pixels on the camera, around 40 × 40 pixels, to capture the total interference patterns for all four output directions. This allows us to still ensure an adequate number of photons per pixel, enabling data acquisition within a manageable time duration. To mitigate the impact of environmental noise in the situation of smaller number of signal photons, we adopt the heralding technique (as shown in Fig. 2d). Subsequently, we develop a fitting procedure for each interference pattern formed at a particular output direction, assuming that only four waves from the involved metalenses interfere with each other. The fitting parameters correspond to the complex amplitudes exiting each metalens. Based on this fitting procedure, we achieve an effective 4x up-sampling of the interference pattern images, which are presented in Fig. 4a and Fig. 4b for the GS and QFT algorithms, respectively. Further details regarding the up-sampling process using the physical fitting model are provided in the Supplementary Materials. The up-sampled interference patterns are displayed in a tabular form with each column labeling the input state and each row labeling the output interference pattern at individual output directions (output basis), resembling to the classical results shown in Fig. 3 by reading out the individual interference patterns in a row-major manner. At the same time, we have also obtained the $\left|\psi_i^{(\text{out})}\right|$ (continuous values now) at each output direction $\boldsymbol{k}_i$ during the fitting process. These values are presented in the same tabular form in Fig. 4c and Fig. 4d for the GS and QFT algorithms, respectively. In both cases, a comparison is made to the expected results (identity matrix), resulting in a root-mean-square error (RMSE) of 0.148 for all 16 elements in the GS algorithm and 0.216 for the QFT algorithm. It is worth noting that the larger error observed at the $k_0$ direction in the QFT algorithm (as shown in Figure 4d) can be attributed to the presence of residual beams, which also appear in Figure 3c, contributing to the larger RMSE in the QFT case.

**Conclusion**

In conclusion, we have developed a metalens array platform capable of implementing quantum algorithms at the single-photon level. The functionality of the platform has been validated using both classical and quantum light sources. By selectively exciting subsets of metalenses and interpreting the interference patterns at specific output directions, we have demonstrated the programmability of quantum algorithms, exemplified by the Grover's search (GS) and Quantum Fourier Transform (QFT) algorithms in this work. Such a platform holds promise for miniaturizing quantum optical circuits and algorithms, opening up possibilities for metasurface applications. Although the scalability to a larger number of qubits through this approach requires further exploration, it already offers the opportunity to 'factorize' certain components within a larger-scale quantum computation. The metasurface can effectively reduce the size and eliminate the need for optical alignment in the factorized part. While we have demonstrated resource sharing between two algorithms on the same metasurface, further optimization of resource management can be explored in future developments. Additionally, the current work utilizes a geometric metasurface with robust operation, relying on metallic components rather than local resonances. However, it may be advantageous to adopt dielectric metasurfaces in the foreseeable future to achieve higher transmission efficiency, thereby enabling the use of larger metalens arrays for quantum operations.

# References

[1] Ladd, T. D. et al. Quantum computers. Nature 464, 45–53 (2010).
[2] C. D. Bruzewicz, J. Chiaverini, R. McConnell, and J. M. Sage, Trapped-Ion Quantum Computing: Progress and Challenges, Applied Physics Reviews 6, 021314 (2019).
[3] H.-L. Huang, D. Wu, D. Fan, and X. Zhu, Superconducting Quantum Computing: A Review, Sci. China Inf. Sci. 63, 180501 (2020).
[4] P.R. Tapster, J.G. Rarity, and P.C.M. Owens, Violation of Bell's Inequality over 4 km of Optical Fiber. Phys. Rev. Lett. **73**, 1923 (1994).
[5] Y.-H. Kim, R. Yu, S. P. Kulik, Y. H. Shih, M. Scully. A Delayed "Choice" Quantum Eraser. Phys. Rev Lett. **84**, 1–5 (2000).
[6] V. Jacques, E. Wu, F. Grosshans, F. Treussart, P. Grangier, A. Aspect, and J.-F. Roch. Experimental realization of Wheeler's delayed-choice gedanken experiment. Science **315**, 966-968 (2007).
[7] H.-S. Zhong et al., Quantum computational advantage using photons, Science, **370**, 1460–1463 (2020).
[8] L. S. Madsen, F. Laudenbach, M. F. Askarani, F. Rortais, T. Vincent, J. FF Bulmer, F. M. Miatto et al. "Quantum computational advantage with a programmable photonic processor." Nature **606**, 75-81 (2022).
[9] Takeda, S. & Furusawa, A. Toward large-scale fault-tolerant universal photonic quantum computing. APL Photonics **4**, 060902 (2019).
[10] Spreeuw, R. J. C. A Classical Analogy of Entanglement. Foundations of Physics 28, 361–374 (1998).
[11] E. Karimi, R. W. Boyd. Classical entanglement. Science **350**, 1172-1173 (2015).
[12] N. J. Cerf, C. Adami, P. G. Kwiat. Optical simulation of quantum logic. Phys. Rev. A, **57**, 1477(R) (1998).

[13] B. Perez-Garcia, J. Francis, M. McLaren, R. I. Hernandez-Aranda, A. Forbes, T. Konrad. Quantum computation with classical light: Implementation of the Deutsch-Jozsa algorithm. Phys. Lett. A, **380**, 1925-1931 (2016).

[14] B. Perez-Garcia, R. I. Hernandez-Aranda, A. Forbes, T. Konrad. The first iteration of Grover's algorithm using classical light with orbital angular momentum. J. Mod. Opt., **65**, 1942-1948 (2018).

[15] S. Zhang, P. Li, B. Wang, Q. Zeng, X. Zhang. Implementation of quantum permutation algorithm with classical light. J. Phys. Commun. **3**, 015008 (2019).

[16] G. Puentes, C. La Mela, S. Ledesma, C. Iemmi, J. P. Paz, M. Saraceno. Optical simulation of quantum algorithms using programmable liquid-crystal displays. Phys. Rev. A, **69**, 042319 (2004).

[17] M. Hor-Meyll, D. S. Tasca, S. P. Walborn, P. H S. Ribeiro, M. M. Santos, and E. I. Duzzioni. Deterministic quantum computation with one photonic qubit. Phys. Rev. A **92**, 012337 (2015).

[18] G. F. Borges, R. D. Baldijão, J. G. L. Condé, J. S. Cabral, B. Marques, M. Terra Cunha, A. Cabello, and S. Pádua. Automated quantum operations in photonic qutrits. Phys. Rev. A **97**, 022301 (2018).

[19] M. Tillmann, B. Dakić, R. Heilmann, S. Nolte, A. Szameit, and P. Walther, Experimental Boson Sampling, Nature Photon **7**, 540 (2013).

[20] F. Yue, D. Wen, J. Xin, B. D. Gerardot, J. Li, and X. Chen, Vector Vortex Beam Generation with a Single Plasmonic Metasurface, ACS Photonics **3**, 1558 (2016).

[21] H.-X. Xu, H. Liu, X. Ling, Y. Sun, and F. Yuan, Broadband Vortex Beam Generation Using Multimode Pancharatnam–Berry Metasurface, IEEE Trans. Antennas Propagat. **65**, 7378 (2017).

[22] Y. Bao, J. Ni, and C. Qiu, A Minimalist Single-Layer Metasurface for Arbitrary and Full Control of Vector Vortex Beams, Adv. Mater. **32**, 1905659 (2020).

[23] G. Zheng, H. Mühlenbernd, M. Kenney, G. Li, T. Zentgraf, and S. Zhang, Metasurface Holograms Reaching 80% Efficiency, Nature Nanotech **10**, 308 (2015).

[24] Z. Zhu et al., Metasurface-Enabled Polarization-Independent LCoS Spatial Light Modulator for 4K Resolution and Beyond, Light Sci Appl **12**, 151 (2023).

[25] K. Wang et al., Quantum Metasurface for Multiphoton Interference and State Reconstruction, Science **361**, 1104 (2018).

[26] T. Stav, A. Faerman, E. Maguid, D. Oren, V. Kleiner, E. Hasman, and M. Segev, Quantum Entanglement of the Spin and Orbital Angular Momentum of Photons Using Metamaterials, Science **361**, 1101 (2018).

[27] L. Li et al., Metalens-Array–Based High-Dimensional and Multiphoton Quantum Source, Science **368**, 1487 (2020).

[28] J. Zhou, S. Liu, H. Qian, Y. Li, H. Luo, S. Wen, Z. Zhou, G. Guo, B. Shi, Z. Liu. Metasurface enabled quantum edge detection. Sci. Adv. 6, eabc4385 (2020).

[29] Li et al., Nat. Photonics 15, 267–271 (2021).

[30] T. K. Yung, J. Xi, H. Liang, K. M. Lau, W. C. Wong, R. S. Tanuwijaya, F. Zhong, H. Liu, W. Y. Tam, and J. Li, Polarization Coincidence Images from Metasurfaces with HOM-Type Interference, IScience **25**, 104155 (2022).

[31] T. K. Yung, H. Liang, J. Xi, W. Y. Tam, and J. Li, Jones-Matrix Imaging Based on Two-Photon Interference, Nanophotonics **12**, 579 (2023).

[32] Q. Li, W. Bao, Z. Nie, Y. Xia, Y. Xue, Y. Wang, S. Yang, and X. Zhang, A Non-Unitary Metasurface Enables Continuous Control of Quantum Photon–Photon Interactions from Bosonic to Fermionic, Nat. Photonics **15**, 267 (2021).

[33] Q. Cheng, T. J. Cui, W. X. Jiang, B. G. Cai. An omnidirectional electromagnetic absorber made of metamaterials. New J. Phys. **12**, 063006 (2010).

[34] K. Cheng, W. Zhang, Z. Wei, Y. Fan, C. Xu, C. Wu, X. Zhang, and H. Li, Simulate Deutsch-Jozsa Algorithm with Metamaterials, Opt. Express **28**, 16230 (2020).

[35] K. Cheng, Y. Fan, W. Zhang, Y. Gong, S. Fei, H. Li. Optical realization of wave-based analog computing with metamaterials. Appl. Sci., **11**, 141 (2020).

[36] V. Bagnoud and J. D. Zuegel, Independent Phase and Amplitude Control of a Laser Beam by Use of a Single-Phase-Only Spatial Light Modulator, Opt. Lett. **29**, 295 (2004).